\documentclass{jfm}

\ifCUPmtlplainloaded \else
  \checkfont{eurm10}
  \iffontfound
    \IfFileExists{upmath.sty}
      {\typeout{^^JFound AMS Euler Roman fonts on the system,
                   using the 'upmath' package.^^J}%
       \usepackage{upmath}}
      {\typeout{^^JFound AMS Euler Roman fonts on the system, but you
                   dont seem to have the}%
       \typeout{'upmath' package installed. JFM.cls can take advantage
                 of these fonts,^^Jif you use 'upmath' package.^^J}%
      }
  \else
  \fi
\fi

\ifCUPmtlplainloaded \else
  \checkfont{msam10}
  \iffontfound
    \IfFileExists{amssymb.sty}
      {\typeout{^^JFound AMS Symbol fonts on the system, using the
                'amssymb' package.^^J}%
       \usepackage{amssymb}%
         
       \let\ge=\geqslant  
      }{}
  \fi
\fi

\ifCUPmtlplainloaded \else
  \IfFileExists{amsbsy.sty}
    {\typeout{^^JFound the 'amsbsy' package on the system, using it.^^J}%
     \usepackage{amsbsy}}
    {\providecommand\boldsymbol[1]{\mbox{\boldmath $##1$}}}
\fi

\newsavebox{\astrutbox}
\sbox{\astrutbox}{\rule[-5pt]{0pt}{20pt}}

\newcommand\p{\ensuremath{\partial}}

\title[Tank-treading propulsion in viscous flows]
{Tank-treading as a means of propulsion in viscous shear flows}
\author[P.~Olla]%
{P\ls I\ls E\ls R\ls O\ns
O\ls L\ls L\ls A$^{1,2}$}

\affiliation{
$^1$ ISAC-CNR, Sez. Cagliari, I--09042 Monserrato, Italy \\[\affilskip]
$^2$ INFN, Sez. Cagliari, I--09042 Monserrato, Italy}

\pubyear{2003}
\volume{???}
\pagerange{1--9}
\date{?? and in revised form ??}
\usepackage{graphicx}
\usepackage{color}

\usepackage{natbib}

\begin{document}

\def\beq{\begin{eqnarray}}
\def\eeq{\end{eqnarray}}
\def\d{{\rm d}}
\def\i{{\rm i}}
\def\e{{\bf e}}
\def\ex{{\rm E}}
\def\bOmega{{\boldsymbol{\Omega}}}
\def\bsigma{{\boldsymbol{\sigma}}}
\def\homega{{\hat\omega}}
\def\r{{\bf r}}
\def\x{{\bf x}}
\def\X{{\bf X}}
\def\bp{{\bar p}}
\def\p{{\bf p}}
\def\g{{\bf g}}
\def\A{{\cal A}}
\def\R{{\bf R}}
\def\u{{\bf u}}
\def\v{{\bf v}}
\def\n{{\bf n}}
\def\f{{\bf f}}
\def\F{{\bf F}}
\def\t{{\bf t}}
\def\im{{\rm i}}
\def\T{{\boldsymbol{\sf T}}}
\def\S{{\boldsymbol{\sf S}}}
\def\I{{\boldsymbol{\sf I}}}
\def\U{{\bf U}}
\def\V{{\bf V}}
\def\Y{{\bf Y}}
\def\cm{{\rm cm}}
\def\l{{\bf l}}
\def\sec{{\rm S}}
\def\Ckol{C_{Kol}}
\def\flux{\bar\epsilon}
\def\Ca{{\rm Ca}}
\def\smali{{\scriptscriptstyle i}}
\def\smalfi{{\scriptscriptstyle \frac{5}{3} }}
\def\smalL{{\scriptscriptstyle{\rm L}}}
\def\smalP{{\scriptscriptstyle {\rm P}}}
\def\smalT{{\scriptscriptstyle {\rm T}}}
\def\smalE{{\scriptscriptstyle{\rm E}}}
\def\smal1n{{\scriptscriptstyle (1.1,n)}}
\def\smaln{{\scriptscriptstyle (n)}}
\def\smalA{{\scriptscriptstyle {\rm A}}}
\def\smalze{{\scriptscriptstyle (0)}}
\def\smalun{{\scriptscriptstyle (1)}}
\def\smaldu{{\scriptscriptstyle (2)}}
\def\smaln{{\scriptscriptstyle (n)}}
\def\smalel{{\scriptscriptstyle (l)}}
\def\smalB{{\scriptscriptstyle B}}
\def\smalT{{\scriptscriptstyle T}}
\def\smalloc{{\scriptscriptstyle ,loc}}
\def\smalglo{{\scriptscriptstyle ,glo}}
\def\gammaP{\gamma^\smalP}
\def\shell{{\rm S}}
\def\ball{{\rm B}}
\def\nav{\bar N}
\def\micron{\mu{\rm M}}
\font\brm=cmr10 at 24truept
\font\bfm=cmbx10 at 15truept

\maketitle

\begin{abstract}
The use of tank-treading as a means of propulsion for microswimmers in viscous
shear flows is taken into exam. We discuss the possibility that a vesicle be able to
control the drift 
in an external shear flow, by varying locally the 
bending rigidity of its own membrane.
% We consider a model microswimmer in the form of a
%a vesicle, with the ability to control the bending rigidity of its own membrane.
By analytical calculation in the quasi-spherical
limit, the stationary shape and the orientation of the tank-treading vesicle in the 
external flow, are determined, working to lowest order in the membrane 
inhomogeneity.
The membrane inhomogeneity acts in the shape evolution equation
as an additional force term, that can be used to balance the effect of the
hydrodynamic stresses, thus allowing the vesicle to assume shapes and orientations that
would otherwise be forbidden. The vesicle
shapes and orientations required for migration transverse to the flow, together with
the bending rigidity profiles  that would lead to such shapes and orientations, are determined.
A simple model is presented, in which a vesicle is able to migrate up
or down the gradient of a concentration field, by stiffening
or softening of its membrane, in response to the variations in the concentration
level experienced during tank-treading.
%membrane elements on its surface, as they tank-tread 
%through points of the fluid where the concentration level is different.
\end{abstract}

\section{Introduction}
\label{sec1}
Microorganisms such as bacteria and protozoa are able to swim in a viscosity
dominated environment through a variety of strategies.
Some of them, such as amoebae and some bacteria, 
exploit deformations in their main body \cite[]{berg76}, others 
utilize cilia \cite[]{blake71,blake74} or flagella \cite[][]{blum79,berg}, 
still others, such as cyanobacteria, are
able to generate travelling waves on their surface \cite[]{ehlers96}.
In all cases, contrary to what happens
at macroscopic scales, fluid inertia plays no role, and microscopic 
swimming is essentially a low Reynolds number affair \cite[][]{lighthill,childress}
(see \cite{lauga09} for a recent review).

One of the motivations for the interest in swimming at low Reynolds numbers
is its relevance for the future realization
of artificial microswimmers, which would have widespread applications
in medicine and in the industry.
Over the years, various propulsion schemes have been proposed,
both discrete (typically,  an assembly of rigid parts hinged together, or connected through 
immaterial links and springs; see e.g. \cite{purcell77,najafi04,avron05}) and continuous 
\cite[]{lighthill52,stone96,ishikawa08}.
In all cases, proper design of a microswimmer entails a complex 
optimization problem, which must take into account limitations,
such as those imposed by the scallop theorem \cite[][]{purcell77,shapere89}.

%Not taking
%care of such limitations, leads to the risk of having a microswimmer
%that executes a large number of swimming strokes without being able to
%move from its initial position.

Recently, progress in mechanical manipulation at the microscale has allowed
to realize the first examples of artificial microscopic swimmers
\cite[][]{dreyfus05,yu06,behkam06,tierno08,leoni09}. 
%This is
%the first step to the construction of ''microbots'' whose application
%would be widespread, e.g. in medicine, as microscopic drug carriers
%in not otherwise accessible regions of the human body. 
At the present
stage, however, most of such artificial swimmers are driven by
external fields and the problem of an autonomous power source 
remains under study. Among the solutions that have been taken into
consideration, various methods of rectification of Brownian motion
\cite[][]{lobaskin08,golestanian09},
and mechanical reactions in the swimmer  
body, induced by inhomogeneity in the environment, e.g.
a chemical gradient \cite[][]{golestanian05,paxton06,pooley07}.

%Now, most theoretical studies of swimming have been carried on hypothesizing a
%quiescent ambient fluid. In a realistic situation, however, a microscopic
%swimmer will move in the presence of external flow gradients due both
%to macroscopic conditions (e.g. the flow inside of a blood vessels) or
%the presence of other swimming microorganisms. These external flow 
%gradients have non-trivial consequences as regards the swimming
%strategies. 

Given the fact that a microswimmer typically lives in a non-quiescent
environment, a possibility that has been taken into consideration, is to exploit 
the velocity fields already present in the fluid as an energy source for
propulsion. Such a swimmer would
sail through the fluid, by a sequence of deformations 
induced in its body by the hydrodynamic stresses in the external flow.
A recent example of such ``passive'' swimming has been illustrated in \cite{olla10},
based on a discrete swimmer design similar to the one considered in 
\cite{najafi04} and \cite{golestanian08}.

It should be mentioned that
passive swimming (at microscopic scales) already exists in nature. An example
is the Fahraeus-Lindqwist effect \cite[]{vand48}: a red cell in a
small artery will deform in response to the flow,
in such a way to be pushed to the vessel center,
thus decreasing its fluid-mechanic resistivity.
In analogous way, vesicles are able to migrate transverse to a wall
bounded shear flow thanks to tank-treading
\cite[][]{olla97,sukumaran01,abkarian02}, and similar 
behavior have been observed in quadratic shear flows as well
\cite[][]{olla00,coupier08,danker09}.

%It should be mentioned that 
Tank-treading
has already been taken into consideration
as a possible microswimmer propulsion system \cite[][]{purcell77,leshansky08}
(see also \cite{tierno08} for a somehow related approach).
It is not too much of a surprise, therefore, that the optimal strategy
for a passive discrete microswimmer in a viscous flow turns out to be
a discrete version of tank-treading \cite[]{olla10}. It is natural 
to ask what would be an appropriate design for a continuous counterpart of this
device. We shall concentrate our analysis on continuous microswimmers
whose basic structure is that of a vesicle.

One of the motivations for the present study is that the efficiency of a discrete swimmer,
of the kind considered in \cite{najafi04} and \cite{golestanian08}, is rather low.
It is in fact $\propto a\delta R/R^2$, where $a$ the size of the moving parts, $\delta R$
is the stroke amplitude, $R$ is the body size; typically: $a/R,\delta R/R\ll 1$.
In the case of a continuous swimmer, instead, $a/R\equiv 1$ and the efficiency would become 
$\propto \delta R/R$. 

\begin{figure}
\begin{center}
\includegraphics[draft=false,width=6.cm]{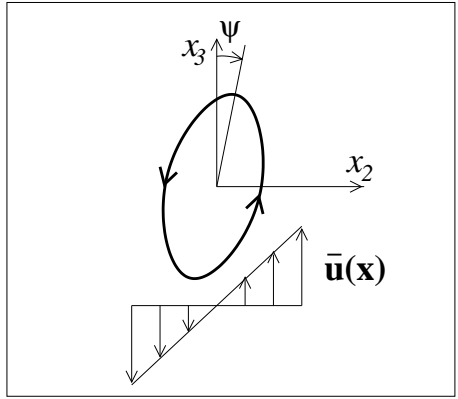}
\caption{
Tank-treading vesicle in a linear shear: the membrane 
circulates around the vesicle interior preserving the shape and orientation of
the object. A vesicle with a homogeneous membrane will maintan an overall ellipsoidal
shape, with long axis at an angle $\psi<\pi/4$ with respect to the direction of the
flow. Low internal viscosities correspond to $\psi\simeq\pi/4$; high internal 
viscosities will lead to $\psi\to 0$ and then to transition to regimes in which the
vesicle rotates more or less as a rigid object.
}
\label{eggfig1}
\end{center}
\end{figure}

The complicated problem lies in the design of an appropriate control 
system for such a device.
In the absence of a control system, a simple vesicle, immersed 
in a linear shear flow,
will stay naturally in a tank-treading condition, provided the viscosity
contrast between interior and exterior fluid is not too high
\cite[][]{kraus96}. A sketch of a tank-treading vesicle in a linear shear 
is provided in Fig. \ref{eggfig1}.
Tank-treading will make such a vesicle migrate away from a solid wall perpendicular to 
the shear plane and parallel to the flow, but, unfortunately, no other migration behaviors 
are possible. Migration towards a wall, for example, would require an 
impossible condition, in which, the tank-treading vesicle maintains an ellipsoidal shape,
with long axis aligned with the contracting direction of the strain component of
the shear. For similar reasons, no transverse migration would be possible in an 
unbounded shear flow. The missing ingredient is some mechanism to generate internal
stresses that counteract the straining action of the external flow.

We want to explore the possibility that a vesicle be able to generate such 
stresses through appropriate modification of its membrane properties, namely,
local stiffening or softening of its outer surface. 
(We hypothesize that the 
energy required for stiffening and softening of the membrane be negligible
compared to the one that would expended to achieve an identical migration 
velocity without the help of the external flow; in this sense, we speak of
a passive swimmer. Although natural at the macroscale, this hypothesis
may require further justification in the case of microscopic objects).
Notice that the presence of a mere inhomogeneity in the membrane, say an inclusion,
would be insufficient to this goal. Such an inhomogeneity would be advected by
the membrane flow and would be unable to lead to a stationary vesicle configuration.
The inhomogeneity of the membrane must not vary with time in a laboratory reference
frame, which implies that the membrane elements must continuously change their 
properties, as they tank-tread around the vesicle.

The mechanism through which local modifications of the membrane stiffness lead
to the generation of stresses in the fluid, closely resembles the one responsible
for the Marangoni effect \cite[]{young59,subramanian}. Marangoni effects have
been considered indeed as a possible mechanism of self-propulsion for droplets
in inhomogeneous environments \cite[]{kitahata02,furtado08}, and \cite{hanna10}
have studied
the Marangoni stresses, generated by an external shear induced redistribution 
of a surfactant over a droplet.

The philosophy in the present paper is different: the membrane stresses
do not act directly, through internal flows in the vesicle, to generate propulsion; 
they rather contribute to modify the vesicle shape, and propulsion is achieved
through interaction between a fixed non-spherical shape and the external flow.

We shall consider the case of an ideal membrane, so that the only material
property that must be taken into account is a bending rigidity.
The specific behavior we shall be interested in is the transverse drift in an 
unbounded linear shear, already considered in \cite{olla10}. We shall determine
the bending rigidity profile that would generate such behavior, and investigate
the possibility that 
the required bending rigidity profiles be obtained as
a direct response of the membrane to the external environment,
without the need of an ``intelligent'', internal control system.

An analytical treatment of the problem is possible only in the case of
quasi-spherical vesicles, and, for this purpose, 
the analysis in \cite{seifert99} and \cite{olla00} will be generalized
to the case of an inhomogeneous membrane.
As discussed in \cite{farutin10}, the perturbative problem is singular and
care must be taken to scale appropriately the shear strength and the viscosity
contrast between inner and outer fluid, in function of the deviation from 
spherical shape of the vesicle. 
For the sake of simplicity, contrary to \cite{seifert99}, all finite temperature 
effect, will be disregarded in the analysis. 

This paper is organized as follows. In Sec. \ref{sec2} the bending forces exerted
on the ambient fluid by a quasi-spherical inextensible membrane, will be calculated,
generalizing to the case of an inhomogeneous bending rigidity, the analysis in 
\cite{zhong-can89}. In Sec. \ref{sec3}, the shape dynamics of a vesicle with an
inhomogeneous membrane, in a viscous shear flow, will be analyzed. In Sec. \ref{sec4}, 
the possibility of drift of a tank-treading vesicle in an unbounded shear flow
will be discussed, and the bending rigidity profiles required for drift will be
determined. In Sec. \ref{sec5}, a simple model of vesicle, with a membrane that
changes properties in response to the external environment, will be presented,
and the migration behavior of the vesicle will be discussed.
Section \ref{sec6} is devoted to conclusions.
Additional technical details will be presented in the Appendices.

%Unless the device is able to store the energy of the flow gradient,
%and to use it at the appropriate moment (say, to
%reach a bacteria), the swimming velocity will be
%rather small, of the order of the velocity
%difference in the external flow at the swimmer scale.
%This may nevertheless be of some interest if one
%is interested in collective effects, in which,
%as in the Fahraeus-Linqwist effect,
%many swimmers move to concentrate in a given region
%of a domain.

%In this paper, the analysis carried on in \cite{olla10} in the case
%of an unbounded flow, is extended to account for the presence
%of a wall, and to determine the structural dynamics in 
%the device that would lead to the desired swimming
%behaviors.  As in \cite{olla10}, the analysis is carried with the simple
%three-sphere model introduced in \cite{najafi04}.
%\\
%This paper is organized as follows. In Sec. \ref{sec2}, the basic 
%equations of the model are presented. In Sec. \ref{sec2.1} the
%results in \cite{olla10} are briefly summarized.
%In Sec. \ref{sec3}, the analysis is extended to the case of
%a wall bounded flow.  In Sec. \ref{sec4},
%the mechanism of energy extraction from the flow is discussed.
%The structural dynamics of the swimmer is discussed in 
%Sec. \ref{sec5}.  Section \ref{sec6} is devoted to conclusions.
%Technical details on the swimmer behavior in a wall bounded flow
%are provided in the Appendix.

\section{The inhomogeneous membrane}
\label{sec2}
The shape of a vesicle can be described in terms of the position $\R(s_1,s_2)$
of the points on the membrane, in function of a suitable set of 
curvilinear coordinates $s_{1,2}$. For a quasi-spherical vesicle, it is natural
to work in spherical coordinates $r,\theta,\phi$, such  that $\R=R(\theta,\phi)\e_r$,
and we write
\beq
R(\theta,\phi)=R_0[1+\tilde R(\theta,\phi)],
\label{shape}
\eeq
where $R_0$ is the radius of the sphere with volume equal to that of the
vesicle.

We can decompose the scalar field $\tilde R$ in 
spherical harmonics: $\tilde R(\theta,\phi)=\sum_{lm}\tilde R_{lm}$ $\times Y_{lm}(\theta,\phi)$.
In analogous way, vector fields, such as e.g.
the displacement $\delta\R(\theta,\phi)$ of a membrane point initially at $\R(\theta,\phi)$,
will be expanded on the vector spherical harmonics basis:
\beq
\Y_{{\rm S}lm}=Y_{lm}\e_r,
\quad
\Y_{{\rm E}lm}=\frac{r\nabla Y_{lm}}{\sqrt{l(l+1)}},
\quad
\Y_{{\rm M}lm}=\e_r\times\Y_{{\rm E}lm},
\label{vector_spherical}
\eeq
so that $\delta\R=\sum_{\mu lm}\delta R_{\mu lm}\Y_{\mu lm}$.
%(the subscripts $\{{\rm S,E,M}\}$ stay for scalar, electric and magnetic and come from the
%origin of this basis as a tool in the study of electromagnetic waves).
The basis in (\ref{vector_spherical}) can easily be verified 
to be orthonormal: $\langle \mu lm|\mu'l'm'\rangle\equiv
\int\Y^*_{\mu lm}\cdot\Y_{\mu'l'm'}\sin\theta\d\theta\d\phi=
\delta_{\mu\mu'}\delta_{ll'}\delta_{mm'}$.

Assuming that the vesicle has volume $V=\frac{4}{3}\pi R_0^3$, it is possible to express
the vesicle area in terms of the components $\tilde R_{lm}$ through the formula:
\beq
S=(4\pi+\epsilon)R_0^2,
\qquad
\epsilon=\frac{1}{2}{\sum_{lm}}'(l^2+l-2)|\tilde R_{lm}|^2+O(\tilde R^3),
\label{area}
\eeq
where $\sum'_{lm}\equiv\sum_{l\ge 2}\sum_{m=-l}^l$ \cite[]{seifert99}. 
The excess area $\epsilon\ll 1$ in Eq. (\ref{area}), which parameterizes the 
deviation from spherical shape, will serve as an expansion parameter for 
the theory.
Notice that the $l=1$ terms in the sum in Eq. (\ref{area}) are identically zero, 
which reflects the fact, 
that the components $\tilde R_{l=1,m}$, to lowest order in $\epsilon$, correspond 
to a rigid displacement of the vesicle. 

%Local inextensibility of the membrane implies a relation between normal ($\delta\R_n$) and
%tangential ($\delta\R_t$) deformation components of the membrane. 
%It is possible to see that local inextensibility leads to the following
%relation between deformation components, valid to lowest order in $\epsilon$:
%\beq
%\delta R_{{\rm E}lm}=\frac{2}{\sqrt{l(l+1)}}\delta R_{{\rm S}lm}.
%\label{inextensibility}
%\eeq
%Notice that, again to lowest order in 
%$\epsilon$: $\n=\e_r$ and $\delta\R_n=\sum_{lm}\delta R_{{\rm S}lm}\Y_{{\rm S}lm}$,
%while $\delta\R_t=\sum_{lm}[\delta R_{{\rm E}lm}\Y_{{\rm E}lm}+
%\delta R_{{\rm M}lm}\Y_{{\rm M}lm}]$.

Following \cite{zhong-can89}, the bending energy of an inhomogeneous membrane 
can be expressed as surface integral
\beq
{\cal H}^B=\frac{1}{2}\int\kappa\ (2H-C)^2\d S,
\label{bending_energy}
\eeq
where $H$ is the mean curvature of the membrane, which can be written in 
the form \cite[]{zhong-can89}: $H=\frac{1}{2}\n\cdot\nabla_\t^2\R$, with $\n$ is
the unit normal
and $\nabla_\t^2$ 
the Laplace-Beltrami operator on the membrane; 
$C$ is called the spontaneous curvature, and $\kappa$ is the 
bending rigidity.  
For the sake of simplicity, we shall assume 
symmetry of the membrane, and set $C=0$.
Following \cite{jenkins77}, 
to enforce inextensibility of the membrane, we include a position dependent 
surface tension in the energy integral
\beq
{\cal H}^B\to{\cal H}={\cal H}^B+\kappa_0\int T\d S,
\label{total_energy}
\eeq
where $T$
plays the role of a Lagrange multiplier coupled to the local area element $\d S$.

Let us suppose the membrane is able to react to the external environment through
local variations of its bending rigidity:
\beq
\kappa=\kappa_0[1+\tilde\kappa(\R,t)],
\label{kappa}
\eeq
where, as in \cite{goulian93}, $\tilde\kappa$ is assumed small and will serve, together with
$\epsilon$, as a basis
for a perturbation expansion in Eqs. (\ref{bending_energy}). 
For the moment, we assume the profile $\tilde\kappa(\R,t)$ to be assigned, 
and postpone to Sec. \ref{sec5} any consideration on the dynamical
mechanisms determining its form.
%As we are not interested in the internal membrane dynamics, 
%we shall assume that the form of the function $\tilde\kappa(\R,t)$ be given,
%possibly in function of 
%external parameters, e.g. the concentration gradient of some chemical 
%in the fluid.

The membrane will act on the fluid with a force density
\beq
\f(\r,t)=-\int\frac{\delta{\cal H}}{\delta\R(\theta,\phi)}\delta(\r-\R(\theta,\phi))\,
\d\theta\d\phi,
\label{force_density}
\eeq
which will be the sum of a bending force $\f^B$ and a tension force $\f^T$.
In the case of a homogeneous membrane, the bending force would be directed along
the normal to the membrane. The space dependence of $\kappa$ produces a tangential
force component. In fact, the variation 
of bending energy produced by a deformation field $\delta\R(\theta\phi)$
can be written in the form
\beq
\delta{\cal H}^B=
\delta{\cal H}^B=\int\delta\R\cdot\Big[\frac{\delta{\cal H}^B}{\delta\R_\n}-
2JH^2\nabla_{\bf t}\kappa\Big]\d\theta\d\phi.
\label{variation_H_B}
\eeq
where $J\d\theta\d\phi=\d S$ is the surface element of the undeformed membrane, and 
subscripts
$\n$ and $\t$ identify normal and tangential vector components. 
The tangential contribution in Eq. (\ref{variation_H_B}) accounts for the variation of
bending rigidity at position $\R$, from tangential displacement of a membrane element 
from position $\R-\delta\R_\t$ to $\R$.

To lowest order in $\epsilon$, $\delta\R_\t$ is a combination of vector harmonics
$\Y_{\mu lm}$ with $\mu={\rm E,M}$. From the relation 
$\int\Y_{{\rm M}lm}\cdot\nabla Y_{l'm'}\d S=0$ [see Eq. (\ref{vector_spherical})],
the tangential component of Eq. (\ref{variation_H_B}) has only
components from $\delta R_{{\rm E}lm}$. As it will soon become clear 
[see Eq. (\ref{inextensibility}) below], this implies that flows on the membrane,
induced by inhomogeneity of $\kappa$, are necessarily associated with vesicle
deformations.
%Thus, from Eq. (\ref{inextensibility}), 
%we do not have to consider membrane flows that are not associated with 
%vesicle deformations.

To explicitly calculate the bending force, we expand 
the mean curvature $H$ and the Jacobian in powers of $\epsilon$.
The mean curvature of a quasi-spherical membrane was calculated in \cite{zhong-can89},
and can be rewritten in the form:
\beq
H=\frac{1}{R_0}[-1+(1+\frac{1}{2}\tilde\nabla^2_\t)\tilde R
-\tilde R(1+\tilde\nabla^2_\t)\tilde R
+O(\epsilon^{3/2})],
\label{curvature}
\eeq
where $\tilde\nabla_\t\equiv R_0\nabla_\t$.
%=\e_\theta\partial_\theta+\frac{\e_\phi}{\sin\theta}\partial_\phi+ O(\epsilon^{1/2})$.
In analogous way, we can write for the Jacobian:
\beq
J=R_0^2\sin\theta\Big[(1+\tilde R)^2+\frac{1}{2}\Big((\partial_\theta\tilde R)^2
+\frac{(\partial_\phi\tilde R)^2}{\sin^2\theta}\Big)+
O(\epsilon^{3/2})\Big].
\label{Jacobian}
\eeq
Substituting into Eq. (\ref{bending_energy}), we obtain the expression for the
bending energy:
${\cal H}^B=\kappa_0\int\{2+2\tilde\kappa[1-\tilde\nabla_{\bf t}^2\tilde R]
+\tilde R\tilde\nabla_{\bf t}^2\tilde R+\frac{1}{2}
(\tilde\nabla_{\bf t}^2\tilde R)^2
+O(\epsilon^{3/2})+O(\tilde\kappa\epsilon)\}\sin\theta\d\theta\d\phi$.
Exploiting Eqs. (\ref{variation_H_B}) and 
(\ref{force_density}), using the expression $\tilde\nabla_\t^2=
\partial_\theta^2+\cot\theta\partial_\theta+(\sin\theta)^{-2}\partial_\phi^2+O(\epsilon^{1/2})$,
and expanding on the basis (\ref{vector_spherical}),
we obtain
the following expression for the bending force density, valid to lowest order in 
$\epsilon$ and $\tilde\kappa$:
\beq
\f^B=&-&\frac{\kappa_0}{R_0^3}\sum_{lm}\{l(l+1)[(l^2+l-2)\tilde R_{lm}
+2\tilde\kappa_{lm})]\Y_{{\rm S}lm}
\nonumber
\\
&-&2\sqrt{l(l+1)}\tilde\kappa_{lm}\Y_{{\rm E}lm}\}\delta(r-R_0).
\label{bending_force}
\eeq
To this order of accuracy, the bending force acts on the fluid at the spherical
surface $r=R_0$. Taking for $\tilde\kappa_{lm}$, the
spectrum produced by a discrete set of inhomogeneities (e.g. inclusions) in the membrane,
Eq. (\ref{bending_force}) would lead to the zero temperature expressions
for the interaction forces among such inhomogeneities,
calculated in \cite{goulian93}. 
Notice that inhomogeneity of the membrane produces a tangential component in the bending force,
that, in order for membrane area to be conserved, must be counterbalanced by tension forces. 

%We will assume that the external hydrodynamic forces are of the same order in $\epsilon$
%as $\f^B$. Thus, also $T=O(\epsilon^{1/2})$ and the expansion for the tension force
%$\f^T$ will begin, as in the case of $\f^B$, with $\f^T$. 

The tension force $\f^T$ 
is obtained from variation of $\int T J\d\theta\d\phi$.
% and must counterbalance the components in the hydrodynamic and bending forces, acting to
%extend the membrane.
We restrict our analysis to a situation in which $T=O(\epsilon^{1/2})$, corresponding
to a weak shear regime in which the hydrodynamic and the bending stresses, that must be
balanced by $\f^T$, are of the
same order [see Eq. (\ref{ordering}) below].
Global area changes are quadratic in $\tilde R$, while local changes are linear, hence,
it is convenient to separate in the tension, global and local contributions: $T=T^{glo}+T^{loc}$.
The variation of the anisotropic part is 
$\int T^{loc}\delta J\d\theta\d\phi
%$=\int T^{loc}[\delta\R\cdot\nabla_\n J+\nabla_\t\cdot(\delta\R_\t J)]\d\theta\d\phi
=\int \delta\R\cdot[2(T^{loc}/R_0)\e_r 
-\nabla_\t T^{loc}+O(\epsilon)]J\d\theta\d\phi$. This leads to the contribution to 
the tension force, to lowest order in $\epsilon$:
$\f^{T,loc}=-(\kappa_0/R_0^3)[2(T^{loc}/R_0)\e_r-\nabla_\t T^{loc}]\delta(r$ $-R_0)$. 

The isotropic contribution to the tension energy is $\kappa_0T^{glo}S$; its variation
is simply $\kappa_0T^{glo}\delta S$, which, from Eq. (\ref{area}) leads immediately to the
result, expanding in vectors spherical harmonics:
$\f^{\smalT\smalglo}=-(\kappa_0T^{glo}/R_0^3)\sum_{lm}(l^2+l-2)\tilde R_{lm}\Y_{{\rm S}lm}
\delta(r-R_0)$ \cite[]{seifert99}. Expanding also $\f^{T,loc}$ in vector spherical harmonics, and
summing to $\f^{\smalT\smalglo}$, we obtain
\beq
\f^{\scriptscriptstyle{T}}\equiv\f^{\smalT\smalglo}+\f^{\smalT\smalloc}=
&-&\frac{\kappa_0}{R_0^3}\sum_{lm}\{[(l^2+l-2)T^{glo}\tilde R_{lm}
+2T^{loc}_{lm}]\Y_{{\rm S}lm}
\nonumber
\\
&-&\sqrt{l(l+1)}T^{loc}_{lm}\Y_{{\rm E}lm}\}
\delta(r-R_0).
\label{tension_force}
\eeq
%and we notice the tangential force component, arising from inhomogeneity of the 
%tension on the membrane.
Setting $\f^B_\t+\f^T_\t=0$, we obtain the local tension $T^{loc}_{lm}$ in the absence
of external flow. Substituting into the normal component $\f^T_\n$, it is possible
to see that the contribution 
by inhomogeneity of the membrane,
to the total normal force $\f_\n=\f^B_\n+\f^T_\n$, is
$\f^{in}_\n=-\frac{2\kappa_0}{R_0^3}\sum_{lm}
(l^2+l-2)\tilde\kappa_{lm}\Y_{{\rm S}lm}$ (we shall identify
contributions by inhomogeneity of the membrane, in general,
with superscript ``$in$''). 
As it could be expected from analogous behavior in the case of 
a homogeneous membrane, the $l=1$ components of $\f^{in}$,  
associated with rigid displacement of the vesicle, do not contribute to the sum. 
Notice however that the $l=1$ components of $\f^B$ and $\f^T$ do not balance in general,
and a vesicle with an inhomogeneous, arbitrarily compressible membrane, could propel
itself in a quiescent fluid through a mechanism analogous to the Marangoni effect.
A combination of inhomoegeneous bending rigidity and surface tension, rather than just an
inhomogeneous surface tension (see e.g. \cite{kitahata02}),
would be responsible in this case for propulsion. 
More precisely, it is $f_{{\rm S}1m}$ that is directly responsible 
for propulsion (we recall that the induced deformation components
$\delta R_{{\rm S}1m}$, to lowest order in $\epsilon$, describe rigid displacements),
while $f_{{\rm E}1m}$ can be shown to generate flows inside the vesicle, analogous
to the convection-like rolls that are present in a droplet experiencing Marangoni 
propulsion.

\section{Deformations in an external shear flow}
\label{sec3}
We want to determine the shape evolution equation for a vesicle with inhomogeneous
membrane, immersed in the shear flow
%Suppose the vesicle is immersed in an external shear flow 
\beq
\bar\u(\x,t)=\alpha x_2\hat\x_3
\label{shear}
\eeq
($x_1=r\sin\theta\cos\phi$, $x_2=r\sin\theta\sin\phi$ and $x_3=r\cos\theta$). 
The derivation closely follows the one in \cite{seifert99}, with additional
care, in light of the results in \cite{farutin10}, given to the singular
behaviors taking place in the limit $\tilde\kappa,\epsilon\to 0$.
Identifying with $\eta_{in}$ and $\eta_{out}$ the dynamical viscosities of the
fluid inside and outside the vesicle, we can introduce dimensionless
costants, the capillary number $ \Ca$ and the viscosity contrast $\lambda$:
\beq
 \Ca=\frac{\eta_{out}\alpha R_0^3}{\kappa_0}
\quad
{\rm and}
\quad
\lambda=\frac{\eta_{in}}{\eta_{out}},
\label{zeta}
\eeq
parameterizing
the relative importance of hydrodynamic forces to internal membrane
stresses, and the ratio between internal and external fluid viscosities.

The
vesicle will produce a flow perturbation $\hat\u$ to be added to $\bar\u$ 
on the outside of the vesicle, and
a flow field $\u$ inside the vesicle. The boundary condition at the membrane
will thus read $\bar\u(\R,t)+\hat\u(\R,t)=\u(\R,t)$, and to this we must add the boundary 
conditions $\hat\u=0$ at $r\to\infty$ and $\u=0$ at $x=0$. 
Let us indicate with capital letters values of the fluid velocity on the membrane. Expanding
in the basis of Eq. (\ref{vector_spherical}), the boundary condition on the membrane 
becomes:
\beq
U_{\mu lm}=\bar U_{\mu lm}+\hat U_{\mu lm}.
\label{boundary_condition}
\eeq
In creeping flow conditions, the viscous forces by the fluid are balanced by the
reaction force exerted by the membrane.
To lowest order in $\epsilon$, the force balance 
at the vesicle surface, is evaluated at $r=R_0$, and the boundary condition equation
(\ref{boundary_condition}) is enforced at $r=R_0$ as well. To this order in $\epsilon$,
from membrane inextensibility, such
boundary conditions are the ones imposed on the fluid by a
rigid spherical surface:
\beq
&&\hat u_{\mu lm}^\smalze(R_0)\equiv\hat U_{\mu lm}^\smalze=
-\bar U_{\mu lm}^\smalze\equiv
-\bar u_{\mu lm}^\smalze(R_0),
\qquad
\mu={\rm S,E};
\nonumber
\\
&&u_{\mu lm}^\smalze(R_0)\equiv U_{\mu lm}^\smalze=
\bar U_{\mu lm}^\smalze\equiv
\bar u_{\mu lm}^\smalze(R_0),
\qquad
\quad\ \
\mu={\rm M},
\label{boundary_condition_0}
\eeq
where superscripts indicate order in $\epsilon^{1/2}$.
In the following, although $\tilde\kappa$ and $\tilde R$ (or $\epsilon^{1/2}$) are not
in general quantities of the same order of magnitude,
we shall use superscripts to indicate simultaneously order 
in $\tilde\kappa$ and $\epsilon^{1/2}$ [for instance, what we have calculated
in Eq. (\ref{bending_force}) is actually 
$\f^{\scriptscriptstyle{B,{\rm (1)}}}$].

We see that $U^\smalze_{{\rm S}lm}=U^\smalze_{{\rm E}lm}
=\hat U_{{\rm M}lm}^\smalze=0$. (Besides, it is possible to see that
absence of
external torques implies $\hat u_{{\rm M}1m}=0$ to all order in $\tilde R$).

To lowest order in $\epsilon$, the force balance equation will take the form:
\beq
f_{\mu lm}+\alpha\eta_{out}[\hat g_{\mu lm}(\hat\U^\smalze)
-g_{\mu lm}(\bar\U^\smalze)+\lambda g_{\mu lm}(\U^\smalun)]\delta(r-R_0)=0,
\label{force_balance}
\eeq
where 
$\hat\g$ and $\g$ indicate components of the adimensionalized hydrodynamic surface force
density,
associated with fluid flow components vanishing, respectively, at $r\to\infty$
and $r\to 0$ (notice that $g$ is defined as as the force density
exerted by a flow inside the membrane, whence the minus sign in front
of the contribution by $\bar\U^\smalze$; see Appendix A). 
Inspection of Eqs. (\ref{bending_force},\ref{force_inner}-\ref{force_outer})
shows us that, in order for all the terms in 
Eq. (\ref{force_balance}) to be of the same order of magnitude, the dimensionless
parameters $ \Ca$ and $\lambda$ must satisfy:
\beq
 \Ca=O(\epsilon^{1/2})
\quad
{\rm and}
\quad
\lambda=O(\epsilon^{-1/2}),
\label{ordering}
\eeq
corresponding to a regime of weak shear and strong viscosity contrast.

The velocity field $\U$ determines the membrane dynamics. 
In particular,
the membrane displacement rate $\dot R(\theta,\phi;t)$ obeys the equation
\cite[]{seifert99}:
\beq
\dot R=U_r+\U\cdot\nabla_\t R.
\label{membrane_displacement}
\eeq
From Eq. (\ref{boundary_condition_0}) [see also Eq. (\ref{shear_component})], 
$\U^\smalze$ is purely due to the vorticity component of $\bar\u$, hence, to
lowest order in $\tilde R$:
$\dot R^\smalun=U_r^\smalun+\bar\u^{rot}_{r=R_0}\cdot\nabla_\t R$,
where 
$\bar\u^{rot}=\frac{1}{2}\alpha(x_2\hat\x_3-x_3\hat\x_2)$, that is the vorticity part
of the shear flow $\bar\u$. The advection term in Eq. (\ref{membrane_displacement}) 
can then be eliminated working in a reference frame rotating with the vorticity of the
flow. We thus have, 
for the $\mu={\rm S}$ component of $\U^\smalun$ needed in 
$g_{\mu lm}^\smalze(\U^\smalun)$:
\beq
U^{r,\smalun}_{{\rm S}lm}=\dot R^r_{lm},
\label{U1}
\eeq
and we have introduced a superscript $r$ as a reminder that the components are calculated
in the rotating reference frame.

It is possible to see that local inextensibility leads to the following
relation between velocity components on the membrane \cite[][]{seifert99,olla00}:
\beq
U^\smalun_{{\rm E}lm}=\frac{2}{\sqrt{l(l+1)}}U^\smalun_{{\rm S}lm}.
\label{inextensibility}
\eeq
Working in the rotating reference frame, Eq. (\ref{inextensibility}) can then be used,
together with Eq. (\ref{U1}), 
to express the components $\U_{{\rm E}lm}$ in Eq. (\ref{force_balance})
in function of $\dot{\tilde R}_{lm}$.

The $\mu={\rm E}$ component of Eq. (\ref{force_balance}) can be used at this point
to express the local tension $T^{loc}_{lm}$ in function of the tangential components
of $\f^B$ and of the hydrodynamic force. Substituting into the $\mu={\rm S}$ component
of Eq. (\ref{force_balance}) and using Eq. 
(\ref{boundary_condition_0}) to express $\hat\U$ in function of $\bar\U$, we remain
with a first order differential equation for the deformation component
$\tilde R_{lm}$. Using 
Eqs. 
(\ref{bending_force}-\ref{tension_force}),
and (\ref{force_inner}-\ref{force_outer}) to explicitate the various force contributions,
we obtain the equation for the deformation dynamics in the rotating reference frame:
\begin{eqnarray}
&&
\lambda \Ca A_l\frac{\d\tilde R^r_{lm}}{\d\tilde t}+
B_l\tilde R^r_{lm}= \Ca\, C^r_{lm}+D_l\tilde\kappa^r_{lm},
\label{olla}
%\\
%&&\frac{1}{2}\sum_{lm}(l^2+l-2)|\tilde R^r_{lm}|^2=\epsilon,
%\label{ollab}
\end{eqnarray}
where 
$\tilde t=\alpha t$, 
\begin{eqnarray}
&&A_l=
%\frac{2l^3+3l^2+4}{l(l+1)}+
\frac{2l^3+3l^2-5}{l(l+1)},
\qquad
%\lambda,
B_l=(l^2+l-2)[l(l+1)+T^{glo}],
\nonumber
\\
&&C^r_{lm}
=
\frac{1}{R_0\alpha}\Big[\frac{4l^3+6l^2-4l-3}{l(l+1)}\bar U^r_{{\rm S}lm}
+\frac{2l+1}{\sqrt{l(l+1)}}\bar U^r_{{\rm E}lm}\Big],
\label{ABCD}
\\
&&D_l=-2(l^2+l-2).
\nonumber
\end{eqnarray}
and the global tension $T^{glo}$ is determined from the constrain equation (\ref{area}).
That all the terms in Eq. (\ref{olla}) contribute to the same order in $\epsilon$
becomes particularly important at the cross-over line in the $\Ca,\lambda$. where
the tank-treading regime [the stationary solution to Eq. (\ref{olla})] is only marginally
stable. As discussed in \cite{farutin10}, analysis of the crossover region for generic
${\rm Ca}$ would require inclusion of all $O(\epsilon)$ terms in Eq. (\ref{olla}).
The choice ${\rm Ca}=O(\epsilon^{1/2})$ allows us to circumvent such difficulties.

For $\tilde\kappa=0$, these equations correspond to the ones obtained in \cite{seifert99} and
\cite{olla00}. They differ only for the expression of the coefficient of the time derivative, that, 
to the order considered in $\epsilon$, due to the ordering in Eq. (\ref{ordering}),
must contain only terms linear in $\lambda$.
%METTERE COMMENTO SU FATTO CHE l=1 NON CONTRIBUISCE A QUEST^{loc}ORDINE MA CONTRIBUIRA`
%A ORDINE SUCCESSIVO NELLA SEZIONE APPROPRIATA.
%%and have introduced dimensionless constants
%$\epsilon$ and $ \Ca$ 
%\beq
%\epsilon=\frac{S}{4\pi R_0^2}-1,
%\qquad
% \Ca=\frac{\eta_{out}\alpha R_0^3}{\kappa_0},
%\eeq
%parameterizing the excess area and the ratio of fluid dynamical and
%elastic time scales.

In order to return to the laboratory frame, it is
sufficient to include in the time derivative in Eq. (\ref{olla}) 
the effect of rotation:
\beq
\frac{\d\tilde R^r_{lm}}{\d\tilde t}\to\frac{\d\tilde R_{lm}}{\d\tilde t}+\sum_{m'}
\Omega_{lmm'}\tilde R_{lm'},
\label{rotating_to_laboratory}
\eeq
where $\Omega_{lmm'}=0$ unless $m'=m\pm 1$, in which case:
\beq
\Omega_{lm,m-1}=\Omega_{lm-1,m}=\frac{\im}{4}\sqrt{(l-m+1)(l+m)}.
\label{Omega}
\eeq
Equation (\ref{olla}) takes then the form in the laboratory frame:
\begin{eqnarray}
\lambda \Ca A_l\Big[\frac{\d\tilde R_{lm}}{\d\tilde t}+\sum_{m'}
\Omega_{lmm'}\tilde R_{lm'}\Big]+B_l\tilde R_{lm}= \Ca\, C_{lm}+
D_l\tilde\kappa_{lm},
\label{olla2}
\end{eqnarray}
where now, from  Eq. (\ref{shear_component}):
\beq
C_{lm}=
2\im\sqrt{\frac{10\pi}{3}}\delta_{l2}\delta_{|m|1}.
\label{C}
\eeq
%Equations  (\ref{olla}) and (\ref{olla2}) are
%essentially the result of a leading order balance
%with respect to $\tilde R$ and $\tilde\kappa$. For this reason
%$O(\lambda^{-1})$ contributions to $A_l$, that were considered
%in [SEIFERT2000] and [OLLA2001] have been disregarded.
%We recall that
%going beyond leading order would be required, if one were
%interested e.g. in the determination to $O(1)$ of the critical
%$\lambda$ for transition out of the tank-treading 
%regime [GRENOBLE].
From Eqs. (\ref{olla}) and (\ref{olla2}), we see that inhomogeneity of the 
membrane acts in the dynamics as a forcing, which acts side by side with the
effect of the external flow. 
Choosing components $\tilde\kappa_{lm}$ appropriately, 
a tank-treading vesicle in an external shear flow
could be stabilized at orientations otherwise 
impossible to achieve (e.g. an ellipsoidal shape with long axis 
along the contracting, rather than the expanding strain direction).

\section{Drift behaviors}
\label{sec4}
Through tank-treading, a vesicle will be able to maintain a fixed shape and 
orientation in a stationary external flow. In the absence of inhomogeneities
in the membrane, a tank-treading vesicle
in the shear flow described by Eq. (\ref{shear}), 
will maintain an ellipsoidal shape with long axis somewhere between the 
stretching direction of the strain and the flow direction $x_3$ \cite[]{kraus96}.
A fixed orientation is the main ingredient allowing migration of a
tank-treading vesicle across the velocity lines of the shear flow, and
a vesicle, in the condition described above, would migrate away from a
solid plane wall perpendicular to the $x_2$ axis.

Non-homogeneity of the membrane provides an additional mechanism to control
the shape and orientation of a vesicle in an external flow, and could be
used in principle to generate drift behaviors that would otherwise be
impossible. We shall focus on the problem of generating a transverse
drift in the flow of Eq. (\ref{shear}), in the case of an unbounded domain.

In order for such a drift to be present, it is necessary that 
the velocity perturbation $\hat\u$
has components $\mu={\rm S,E}$, $l=1$, signaling the presence of a net hydrodynamic
force acting on the vesicle (see Appendix A). To obtain such harmonics, we must include
in the boundary condition
Eq.  (\ref{boundary_condition}) determining $\hat\u$,
the effect of non-sphericity of the surface $r=R(\theta,\phi)$. 
The procedure parallels the one in \cite{olla00}.
To $O(\epsilon^{1/2})$, we can write:
\beq
\hat\U^\smalun=-(R-R_0)\frac{\partial}{\partial r}(\bar\u+\hat\u^\smalze)_{r=R_0}
+\U^\smalun,
\label{boundary_condition_1}
\eeq
from which we get the boundary condition
$\hat\u^\smalun_{r=R_0}=\hat\U^\smalun$.

Passing to vector spherical harmonics, Eq. (\ref{boundary_condition_1}) will take
the form:
\beq
\hat U^\smalun_{\mu lm}=-\sum_{\mu'l'm'}\langle\mu lm|R|\mu'l'm'\rangle U'_{\mu'l'm'}
+U^\smalun_{\mu lm},
\label{temp1}
\eeq
where $\U'\equiv
\frac{\partial}{\partial r}(\bar\u^\smalze+\hat\u^\smalze)_{r=R_0}$.
Exploiting Eqs. (\ref{inner}-\ref{shear_component}) and (\ref{boundary_condition_0}), 
we can write:
\beq
U'_{{\rm S}lm}=0;
\qquad
U'_{{\rm E}lm}=\frac{2l+1}{R_0}\Big(\frac{-3\bar U^\smalze_{{\rm S}lm}}{\sqrt{l(l+1)}}+
2\bar U^\smalze_{{\rm E}lm}\Big)=
\im\sqrt{5\pi}\alpha\delta_{l2}\delta_{|m|1}.
\label{U'}
\eeq
It is possible to see that the contribution to drift from $U^\smalun_{\mu lm}$ 
vanishes identically. In fact, in the rotating
reference frame,   the components $U^\smalun_{\mu lm}$ , $\mu={\rm S,E}$ are related to
$\dot R_{lm}$ through Eqs. (\ref{U1}) and (\ref{inextensibility}). From Eq. 
(\ref{rotating_to_laboratory}), we find in the laboratory frame:
\beq
U^\smalun_{{\rm E}lm}=\frac{2}{\sqrt{l(l+1)}}U^\smalun_{{\rm S}lm}=
\frac{2}{\sqrt{l(l+1)}}\Big(\dot R_{lm}+\sum_{m'}\Omega_{lmm'}R_{lm'}\Big),
\nonumber
\eeq
and we see immediately that $U^\smalun_{\mu lm}=0$ for $\mu={\rm S,E}$; $l=1$.

Returning to Eq. (\ref{temp1}), we see from Eqs. (\ref{U'}) and
(\ref{cacca}) that the only surviving terms in the sum are those for $\mu=\mu'={\rm E}$,
$l=m=1$ and $l'=2$, $m'=\pm 1$. Expanding $\langle\mu lm|R|\mu'l'm'\rangle=\sum_{l''m''}
\langle\mu lm|Y_{l''m''}|\mu'l'm'\rangle R_{l''m''}$, the sum to right
hand side of Eq. (\ref{temp1}) reduces essentially to two terms, involving
matrix elements:
\beq
\langle 11|Y_{30}|21\rangle=\frac{1}{4}\sqrt{\frac{7}{15\pi}},
\qquad
\langle 11|Y_{32}|2,-1\rangle=\frac{1}{\sqrt{14\pi}}.
\nonumber
\eeq
Substituting, together with Eq. (\ref{U'}), into (\ref{temp1}) and then into 
Eqs. (\ref{cacca}-\ref{drift}), we finally obtain:
\beq
\frac{U^{drift}_1}{\alpha R_0}=\sqrt{\frac{5}{21\pi}}Im[\tilde R_{32}],
\quad
\frac{U^{drift}_2}{\alpha R_0}=\sqrt{\frac{5}{21\pi}}
\Big(\frac{7}{4}\sqrt{\frac{2}{15}}\tilde R_{30}+Re[\tilde R_{32}]\Big).
\label{drift_magnitude}
\eeq
A similar coupling between the $l=2$ (or $l=3$) harmonics in a shear flow and $l=3$ 
(or $l=2$) harmonics
in the internal properties of a droplet immersed in the flow, 
has been shown in  \cite{hanna10} to
induce transverse migration of the droplet.

\begin{figure}
\begin{center}
\includegraphics[draft=false,width=10.cm]{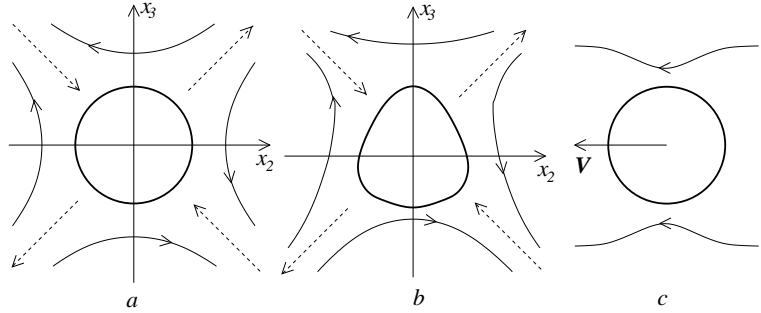}
\caption{Velocity perturbation ($\hat\u$ field) around particles in a viscous fluid. 
$(a)$: a sphere suspended in the shear flow $\bar\u=\alpha x_2\hat\x_3$ ($\alpha>0$; 
dashed arrows indicate the strain component of $\bar\u$); $(b)$:
a tank-treading vesicle suspended in the same flow, shaped to drift towards 
$x_2>0$ [see Eq. (\ref{drift_magnitude})]; $(c)$: a sphere pulled with velocity $\V$ to the
left. In case $(a)$, $\hat u_2(x_3)=-\hat u_2(-x_3)$. In case $(b)$,
$\hat u_2(x_3)\ne-\hat u_2(-x_3)$, and
$\hat u_2(x_3)$ has
an even component which can be shown to have the same sign as the
corresponding one in the velocity dipole in $(c)$. In both cases $(b)$ and $(c)$, the
net effect is a hydrodynamic force pushing the particle to the right.
}
\label{eggfig2}
\end{center}
\end{figure}

From $Y_{30}\propto\cos\theta(5\cos^2\theta-3)$ and 
$Y_{32}\propto\ex^{2\im\phi}\sin^2\theta\cos\theta$, we see that a tank-treading
vesicle drifting to positive $x_2$ will need to have a shape, whose section in 
the shear plane $x_2x_3$ (for $\alpha>0$) is an egg with the tip 
at $x_3>0$. 
The geometrical mechanism for drift along $x_2$ is illustrated in Fig. \ref{eggfig2}
and parallels what is obtained in the case of the discrete swimmer discussed in \cite{olla10}.
Drift to positive $x_1$ will require, on the other hand
a shape whose
section in the $x_1x_2$ plane is an ellipse with the long axis at $\phi=\pi/4$ 
with respect to $x_1$. A discrete version of a passive swimmer undergoing such
a kind of chiral migration has been illustrated in \cite{watari09}.
In both \cite{olla10} and \cite{watari09}, the drift was generated
in an ensemble of connected spheres rotating in a shear flow,
imposing a configuration that was on the average asymmetric 
in the laboratory reference frame.

We can imagine at this point a hypothetical microswimmer, whose structure is that of
a vesicle, with full control of the mechanical properties of its membrane, and
ask what modification of $\kappa$ would be required to achieve the
drift behaviors described in Eq. (\ref{drift_magnitude}). 

From Eq. (\ref{drift_magnitude}), we see that the drift is maximized if all the
excess area is stored in the deformation components $\tilde R_{3,\pm 2}$.
In order for $\tilde R_{2,\pm 1}=0$,
we need that the forcing from the strain components of $\bar\u$ in 
Eq. (\ref{olla2}), be canceled by the contribution by membrane inhomogeneity.
From Eq. (\ref{C}): 
\beq
\tilde\kappa_{2,\pm 1}=\frac{\im}{4}\sqrt{\frac{10\pi}{3}} \Ca,
\eeq
The $l=3$ components of $\tilde\kappa$ are obtained 
imposing 
in Eq. (\ref{olla2})
the tank-treading condition
$\partial\tilde R_{3,\pm 2}/\partial\tilde t=0$, together with 
$\tilde R_{3m}=0$ for $m\ne \pm 2$.
Using Eqs. (\ref{ABCD}) and (\ref{Omega}):
\beq
&&\tilde\kappa_{3,\pm 3}=-\frac{19\sqrt{6}}{240}\im \Ca\,\lambda\,\tilde R_{3,\pm 2},
\quad
\tilde\kappa_{3,\pm 1}=-\frac{19\sqrt{10}}{240}\im \Ca\,\lambda\,\tilde R_{3,\pm 2},
\nonumber
\\
&&\tilde\kappa_{3,\pm 2}=-10(12+T^{glo})\tilde R_{3,\pm 2}.
\label{kappas}
\eeq
We see from  the third in Eq. (\ref{kappas}) that the component $\tilde\kappa_{3,\pm 2}$ 
depends on tension $T^{glo}$, which remains undetermined. A lower bound for the amplitude
$|\tilde\kappa_{3,\pm 2}|^2$ can be obtained requiring stability 
of the configuration, i.e. $B_l>0$, for which it is sufficient that $B_2>0$,
i.e. $T^{glo}>-6$.
From Eq. (\ref{area}),
we find therefore: 
\beq
|\tilde\kappa_{32}|^2>(9/5)\epsilon,
\label{bound}
\eeq
and the arbitrariness of $\tilde\kappa_{32}$ 
reflects the independence of the two expansion parameters of the theory 
$\epsilon^{1/2}$ and $\tilde\kappa$.

%\section{A more realistic example}
\section{A vesicle that ``swims'' in response to concentration gradients}
\label{sec5}
A microswimmer, such as the one described in the previous section, would probably
require a sophisicated control system to achieve the bending rigidity modifications
described in Eqs. (\ref{kappas}-\ref{bound}). 
One may ask whether 
a simpler design is possible, in which the membrane reacts directly to the external
environment, without the need of an internal control system.

We are going to describe such a design, in which the vesicle is able to migrate up (or
down) a concentration gradient in the shear plane, through softening (or stiffening)
of the membrane, in response to a local property of the fluid, such as e.g. the presence
of a chemical substance, a temperature inhomogeneity, 
or a light intensity gradient. 
A key ingredient will appear to be the presence of a delay in the
membrane response to the external environment. 
%Of course, the design is again hypothetical, 
%in that this author is not aware of any material that would at present exhibit such a 
%behavior.

Let us assume that the response of a membrane element in our vesicle, 
to a concentration field $\Theta$,
be described by a linear relaxation equation in the form:
\beq
(\partial_t+\U\cdot\nabla_\t+\gamma)\tilde\kappa=\beta\Theta,
\label{stiffening}
\eeq
where $\gamma$ and $\beta$ could be in general isotropic operators.
%and all the field in the equation are evaluated on the membrane. 
Equation (\ref{stiffening}) could be seen as the result, say, of a process
of absorption or chemical reaction with the environment (the field $\Theta$ may
describe e.g. an absorption flux from the bulk, $\beta$ would be an absorption 
constant and $\gamma\kappa\propto -\nabla_\t^2\kappa$ may account for surface
diffusion effects).
The flow on the tank-treading membrane is
accounted for by the 
advection term $\U\cdot\nabla_\t\tilde\kappa$.
In stationary conditions,
the time derivative will drop off Eq. (\ref{stiffening}).

Let us assume the presence of a concentration gradient along $x_2$, so that,
on the membrane:
$\Theta=\Theta(\R)=\Theta_0+\Theta'R(\theta,\phi)\sin\theta\sin\phi$. 
(We assume that the diffusive current
responsible for the gradient in $\Theta$ is much larger than its advective counterpart
$\sim\alpha\Theta'R$, generated by the flow perturbation due to the vesicle).
The effect of the constant part $\Theta_0$ is an isotropic contribution to $\tilde\kappa$
that could be reabsorbed in a renormalization of $\kappa_0$.
% and it may be necessary to
%invoke some non-linear
%saturation mechanism acting selectively on the $l=0$ component of $\tilde\kappa$
To determine the 
anisotropic part, we expand Eq. (\ref{stiffening}) perturbatively in $\epsilon$:
\beq
(\U^\smalze\cdot\nabla_\t+\gamma)\tilde\kappa^\smalun&=&\Theta'R_0\beta 
\sin\theta\sin\phi,
\nonumber
\\
(\U^\smalze\cdot\nabla_\t+\gamma)\tilde\kappa^\smaldu&=&\Theta'R_0\beta 
\tilde R^\smalun
\sin\theta\sin\phi-\U^\smalun\cdot\nabla_\t\tilde\kappa^\smalun,
\label{stiffening_perturbative}
\eeq
and so on to higher orders.

Let us focus on the regime in which the relaxation time scale for the membrane properties
is much longer 
%(at least for the inhomogeneous part)
than that of the hydrodynamics, that is $\alpha^{-1}$.
In this regime, a membrane will soften (or stiffen) while cruising at $x_2>0$, 
and start stiffening (or softening) when crossing to $x_2<0$. In the velocity field described
in Eq. (\ref{shear}),
therefore, our vesicle will present a softer (stiffer) side to $x_3>0$, and we would expect
an egg shape with tip pointing at $x_3>0$ (at $x_3<0$). From Eq. (\ref{drift_magnitude}),
this would correspond to drift to positive (negative) $x_2$.
Unfortunately,
a linear theory, based only on the first of Eq. (\ref{stiffening_perturbative}),
turns out be insufficient to account for this effect.

%Things, however, are not that simple.

Proceeding as before, we expand Eq. (\ref{stiffening_perturbative}) in spherical harmonics.
The lowest order contribution to advection reads
$\U^\smalze\cdot\nabla_\t= \bar\u^{rot}_{r=R_0}\cdot\nabla$; 
using Eq. (\ref{rotating_to_laboratory}), the first of Eq. (\ref{stiffening_perturbative})
becomes:
\beq
\alpha\sum_{m'}\Omega_{lmm'}\tilde\kappa^\smalun_{lm'}+\gamma_l\tilde\kappa^\smalun_{lm}
=\beta_l\Theta'R_0\langle lm|\sin\theta\sin\phi\rangle,
\label{stiffening_perturbative_1}
\eeq
where now $\gamma_l$ and $\beta_l$ are numbers. We see immediately that $\tilde\kappa^\smalun$
is a superposition of $l=1$ harmonics, which do not contribute, to lowest order in $\epsilon$, 
to the vesicle dynamics [see Eqs. (\ref{olla},\ref{ABCD},\ref{olla2})].
Thus, to lowest order in $\tilde\kappa$, 
the shape of a vesicle in the shear flow of Eq. (\ref{shear})
will be the same as in the case of a homogeneous membrane: an ellipsoid with 
the long axis between the stretching direction of the flow and the $x_3$ axis.

At this point, two strategies are possible: one is 
to replace Eq. (\ref{stiffening}) by
a nonlinear model equation; the other is to 
take into account higher order terms in $\epsilon$ in
the vesicle response to $\Theta$. Now, an equation like (\ref{stiffening}) describes 
the response of the membrane to the small variations of $\Theta$ that occur on the
scale of the vesicle. The physical meaning of a nonlinear version of such equation
remains therefore unclear. On the other hand, the higher order terms in the response
to $\Theta$, could provide qualitative information on the behavior of strongly 
non-spherical vesicles. This suggests us to opt for the second strategy, and to focus on the 
higher order contributions to the membrane response.

%Going to higher order in $\epsilon$, 
We must consider the secondary deformations induced
by $\tilde\kappa^\smaldu$, and by those non-linear contributions to the 
$\tilde\kappa$-dependent part of the force exerted 
by the membrane on the fluid, that were disregarded in Eqs. (\ref{bending_force}) 
and (\ref{tension_force}). Writing in explicit form:
\beq
\f^{in}=\f^{in,L}+\f^{in,N},
\label{total_force_inhomogeneous}
\eeq
with $\f^{in,L}$ identifying the contribution by $\tilde\kappa^\smaldu$, 
$\f^{in,N}$ accounting for the nonlinear part of the force.

Notice that 
the force terms in Eq. (\ref{total_force_inhomogeneous}) are $O(\tilde\kappa\epsilon^{1/2})$.
To the same order of accuracy, also $O(\epsilon)$ terms should be taken into account;
one example are the corrections from approximating $R=R_0$ in the Dirac deltas entering
$\f^B$ and $\f^T$ [see  Eqs. (\ref{bending_force}) and (\ref{tension_force})].
However, from symmetry of the flow, terms that do not involve $\tilde\kappa$ are
superpositions of even $l$ harmonics, while drift is produced by $l=3$ harmonics
[see Eq. (\ref{drift_magnitude})]. Such $O(\epsilon)$ contributions
to the force can thus be disregarded.

\subsection{Higher order contributions to the bending rigidity}
Let us consider first the contribution to the membrane force from $\tilde\kappa^\smaldu$, 
i.e. the term $\f^{in,L}$ in Eq. (\ref{total_force_inhomogeneous}).
First, however, we have to evaluate the components $\tilde\kappa_{lm}^\smalun$.
From $\langle 1,\pm 1|\sin\theta\sin\phi\rangle$ $=\im\sqrt{2\pi/3}$, and
using Eq. (\ref{Omega}) in Eq. (\ref{stiffening_perturbative_1}), 
we find, in the limit $\gamma_1/\alpha\to 0$:
\beq
\tilde\kappa^\smalun_{lm}=
4\sqrt{\frac{\pi}{3}}\hat\kappa\delta_{l1}\delta_{m0},
\qquad
\hat\kappa=\frac{\beta_lR_0\Theta'}{\alpha}.
\label{kappa1}
\eeq
In order to determine $\tilde\kappa^\smaldu$, we have to solve the second of Eq. 
(\ref{stiffening_perturbative}). From Eqs.  (\ref{U1},\ref{inextensibility}) and
(\ref{rotating_to_laboratory}), we have 
$\U^\smalun\cdot\nabla_\t=\alpha\sum_{lmm'}\frac{2}{\sqrt{l(l+1)}}\Omega_{lmm'}R^\smalun_{lm'}
\Y_{{\rm E}lm}\cdot\nabla$.
From Eq. (\ref{C}) and
the fact that $\tilde\kappa^\smalun$ 
has only $l=1$ components, $R^\smalun$ will 
be a superposition of $l=2$ components.
Substituting into the second of Eq. 
(\ref{stiffening_perturbative}), passing to spherical harmonics,
and using Eq. (\ref{kappa1}), we obtain therefore:
\beq
\tilde\gamma_l\tilde\kappa^\smaldu_{lm}+ \Ca\,\lambda
\sum_{m'}\Omega_{lmm'}\tilde\kappa^\smaldu_{lm'}
&=&\hat\kappa\Big[\sum_{m'}\langle lm|Y_{2m'}\sin\theta\sin\phi\rangle\tilde R^\smalun_{2m'}
\nonumber
\\
&-&\frac{8}{3}\sqrt{\frac{2}{3}}\sum_{m'm''}\Omega_{2m'm''}\langle lm|\Y_{{\rm E}lm'}\cdot
\tilde\nabla Y_{10}\rangle\tilde R^\smalun_{2m''}\Big],
\label{stiffening_perturbative_2}
\eeq
where $\tilde\gamma_l=\gamma_l/\alpha$. 

We shall need only the $l=3$ components of $\tilde\kappa^\smaldu$.
Solution of Eq. (\ref{stiffening_perturbative_2}), 
using Eq. 
(\ref{Omega}) and the expressions for the matrix elements provided in Appendix C, 
gives then the result, for $\tilde\gamma_3\to 0$:
\beq
&&\tilde\kappa^\smaldu_{30}
=\frac{2\hat\kappa}{3\sqrt{35}}\Big(11\tilde R^\smalun_{20}+2\sqrt{\frac{2}{3}}
\tilde R^\smalun_{22}\Big),
\nonumber
\\
&&\tilde\kappa_{31}^\smaldu=\sqrt{\frac{10}{7}}\hat\kappa\tilde R^\smalun_{21},
\quad
\tilde\kappa_{32}^\smaldu=\frac{2\hat\kappa}{\sqrt{7}}\tilde R^\smalun_{22},
\quad
\tilde\kappa_{33}^\smaldu=-\frac{\hat\kappa}{3}\sqrt{\frac{2}{21}}\tilde R^\smalun_{21}.
\label{kappa2}
\eeq
The contribution to the bending force is in the same form as Eq. (\ref{bending_force}):
\beq
\f^{B,in,L}
=-\frac{2\kappa_0}{R_0^3}\sum_{lm}\tilde\kappa^\smaldu_{lm}
[l(l+1)\Y_{{\rm S}lm}-\sqrt{l(l+1)}\Y_{{\rm E}lm}]\delta(r-R_0).
\nonumber
\eeq
To this, we must add a tension force contribution $\f^{T,in,L}$, whose effect, as in 
the derivation of Eq. (\ref{olla}), is to cancel
the tangential part of $\f^{B,in,L}$. Using Eq. (\ref{tension_force}), we obtain
\beq
f^{in,L}_{{\rm S}lm}=-\frac{2\kappa_0}{R_0^3}(l^2+l-2)\tilde\kappa^\smaldu_{lm}\delta(r-R_0),
\label{f_in_L}
\eeq
which will lead to a correction 
term $D_l\tilde\kappa^\smaldu_{lm}$ in Eqs. (\ref{olla}) and (\ref{olla2}).

\subsection{Nonlinear corrections in the bending force}
To lowest order in $\tilde R$, the contribution to force by the $l=1$ components in 
$\tilde\kappa$ is identically zero. We must consider terms $\propto\tilde\kappa\tilde R$
in the bending force, which requires keeping terms $\propto\tilde\kappa\tilde R^2$ in 
the bending energy. The lowest order contribution to the bending energy due to the
component $\tilde\kappa^\smalun$ is obtained from Eqs. 
(\ref{bending_energy},\ref{curvature},\ref{Jacobian}):
\beq
{\cal H}^{B,in,N}
&=&\kappa_0\int\Big\{\tilde\kappa^\smalun\Big[\tilde R^\smalun\tilde\nabla_\t^2\tilde R^\smalun+
\frac{1}{2}(\tilde\nabla_\t^2\tilde R^\smalun)^2\Big]
\nonumber
\\
&-&\tilde R^\smalun\Big[
(\partial_\theta\tilde R^\smalun)(\partial_\theta\tilde\kappa^\smalun)
+\frac{(\partial_\phi\tilde R^\smalun)(\partial_\phi\tilde\kappa^\smalun)}{\sin^2\theta}\Big]\Big\}
\sin\theta\d\theta\d\phi.
\nonumber
\eeq
Using Eq. (\ref{force_density}), this corresponds to a force density:
\beq
\f^{B,in,N}&=&\frac{\kappa_0}{R_0^3}
\Big\{[2\tilde\kappa^\smalun(1+\tilde\nabla_\t^2)\tilde\nabla_\t^2\tilde R^\smalun
+(\tilde\nabla^2_\t\tilde\kappa^\smalun)(\tilde\nabla_\t^2\tilde R^\smalun)
\nonumber
\\
&+&\frac{1}{2}\tilde R^\smalun\tilde\nabla^4_\t\tilde\kappa^\smalun]\e_r
+2(\tilde\nabla^2_\t\tilde R^\smalun)\tilde\nabla_\t\tilde\kappa^\smalun
\Big\}\delta(r-R_0).
\label{tmp1}
\eeq
As with the other force terms, a tension contribution must be added, that cancels
the tangential component in Eq. (\ref{tmp1}) and produces a correction to the
normal part, with $f^{T,in,N}_{{\rm S}lm}=\frac{\sqrt{l(l+1}}{2}f^{T,in,N}_{{\rm E}lm}$
[see Eq. (\ref{tension_force})]. In terms of vector spherical harmonics, the resulting
total force will read, using Eq. (\ref{kappa1}):
\beq
f^{in,N}_{{\rm S}lm}=&-&\frac{8\kappa_0}{R_0^3}\sqrt{\frac{\pi}{3}}
\Big\{37\langle lm|Y_{10}|2m\rangle
\nonumber
\\
&-&\frac{12}{\sqrt{l(l+1)}}
\langle{\rm E}lm|\tilde\nabla_\t Y_{10}|2m\rangle\Big\}\hat\kappa\tilde R^\smalun_{2m}
\delta(r-R_0),
\label{f_in_N}
\eeq
and again, to determine the drift, only the $l=3$ force components will be required.

\subsection{Contribution to drift}
The $l=3$ components of the total membrane force are obtained 
putting together Eqs. (\ref{kappa2},\ref{f_in_L},\ref{f_in_N}).
Substituting into Eq. (\ref{olla2}) and using the expression for the matrix elements
provided in Appendix C, leads to the following equation for the $l=3$ component of the 
secondary vesicle deformation:
\beq
\lambda\Ca A_3\sum_{m'}\Omega_{3mm'}\tilde R^\smaldu_{3m'}+B_3\tilde R^\smaldu_{3m}
=\hat\kappa\sum_{m'}E_{3mm'}R^\smalun_{2m'}
\label{olla3}
\eeq
where the only non-zero entries of the matrix $E_{3mm'}$ are:
\beq
&&E_{300}=-\frac{1628}{3\sqrt{35}},
\quad
E_{302}=-\frac{80}{3}\sqrt{\frac{6}{35}},
\nonumber
\\
&&E_{311}=-364\sqrt{\frac{2}{35}},
\quad
E_{322}=-\frac{172}{\sqrt{7}},
\quad
E_{331}=-\frac{20}{3}\sqrt{\frac{2}{21}}.
\label{E}
\eeq
The lowest order components $\tilde R^\smalun_{lm}$ describe the shape of a vesicle
with homogeneous membrane, in the flow (\ref{shear}). The behavior in a viscous shear
flow of a vesicle with such characteristics is well understood 
\cite[][]{noguchi07,lebedev07,farutin10}; the features relevant to the present analysis
are summarized in Appendix B. We have:
\beq
\tilde R^\smalun_{22}=\frac{1}{\sqrt{6}}\tilde R^\smalun_{20}=
\frac{1}{4}\frac{\lambda}{\lambda_{cr}}\epsilon^{1/2},
\qquad
\tilde R^\smalun_{21}=\frac{\im}{2}
\sqrt{1-(\lambda/\lambda_{cr})^2}\ \epsilon^{1/2},
\label{R1}
\eeq
where 
\beq
\lambda_{cr}=4\sqrt{\frac{10\pi}{3\epsilon}}
\label{lambda_cr}
\eeq
is the maximum viscosity contrast for which tank-treading is possible.
In the small $\Ca$ regime considered [see Eq. (\ref{ordering})], for $\lambda>\lambda_{cr}$,
the vesicle will make direct transition to a tumbling regime, in which the vesicle
rotates in the shear plane as a rigid object \cite[]{farutin10}.

The membrane tension, entering the relaxation coefficient $B_3$ in 
Eq. (\ref{olla3}) [see the second of Eq. (\ref{ABCD})], is determined,
to lowest order in $\epsilon$ and $\hat\kappa$, by the shape dynamics
accounted for by Eq. (\ref{R1}):
\beq
T^{glo}=-6+\frac{\lambda_{cr} \Ca}{4}\sqrt{1-(\lambda/\lambda_{cr})^2}.
\label{barT}
\eeq
To the order considered in Eq. (\ref{olla3}), the contribution to tension 
by the secondary deformation $\tilde R^\smaldu$ is disregarded.

\begin{figure}
\begin{center}
\includegraphics[draft=false,width=8.cm]{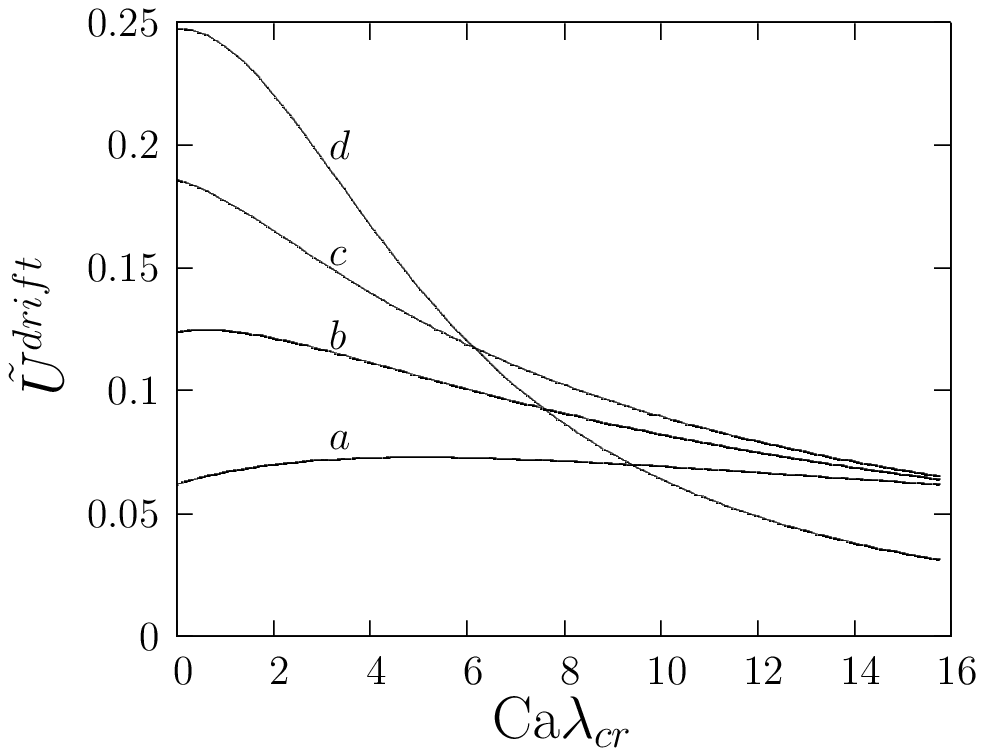}
\caption{Profiles of the normalized drift velocity
$\tilde U^{drift}=-U^{drift}_2/(\alpha R_0\hat\kappa\epsilon^{1/2})$,
in function of $ \Ca\,\lambda_{cr}$ for different values of $\lambda/\lambda_{cr}$. In 
the four cases:
$(a)$: $\lambda/\lambda_{cr}=0.25$; ($b$): $\lambda/\lambda_{cr}=0.5$; ($c$):
$\lambda/\lambda_{cr}=0.75$; ($d$): $\lambda/\lambda_{cr}=1$.
The drift vanishes in the limit $\lambda/\lambda_{cr}\to 0$.
}
\label{eggfig3}
\end{center}
\end{figure}

Using Eqs. (\ref{E}-\ref{barT}), Eq. (\ref{olla3}) can be solved in 
terms of the dimensionless parameters $\hat\kappa$, $\lambda\Ca$ and
$\lambda_{cr}$ (or $\epsilon$). From here, substituting into Eq. (\ref{drift_magnitude}),
the drift velocity can be expressed in the form 
\beq
U^{drift}_2=-\tilde U^{drift}(\lambda/\lambda_{cr}, \Ca\,\lambda_{cr})\hat\kappa\epsilon^{1/2}
\alpha R_0.
\nonumber
\eeq 
Notice that the arguments of $\tilde U^{drift}$, given the scaling in Eq. 
(\ref{ordering}), are not singular in the limit $\epsilon\to 0$.
The profile of the normalized drift velocity $\tilde U^{drift}$ is illustrated 
in Fig. \ref{eggfig3}; we see that the maximum is attained
for $\lambda/\lambda_{cr}=1$ and $\Ca\,\lambda_{cr}=0$, in which case
\beq
U^{drift}_2=-\frac{4279}{37800}\sqrt{\frac{15}{\pi}}\hat\kappa\epsilon^{1/2}\alpha R_0\simeq
-0.25\hat\kappa\epsilon^{1/2}\alpha R_0.
\label{Udrift}
\eeq
As expected, softening of the membrane in regions of higher $\Theta$,
(which implies $\hat\kappa<0$), will lead to an up-gradient drift of the vesicle.

The increase of $\tilde U^{drift}$ as $\lambda/\lambda_{cr}\to 1$ and $\Ca\lambda_{cr}\to 0$,
shown in Fig. \ref{eggfig3},
is associated with a complex pattern of vesicle deformations.
Passing from $\lambda=\lambda_{cr}$ to $\lambda=0$ for $\Ca\,\lambda_{cr}$ fixed, 
the long axis of the ellipsoid described
by the components $R^\smalun_{2m}$ will shift from alignment with the flow, to an
orientation at $\pi/4$ with respect to it. At the same time, 
the components $R^\smaldu_{3m}$ with $m=0,2$, that are associated
with drift, will go to zero in the limit. [This is consequence of the fact that
the matrix elements $E_{lmm}$, as described in Eq. (\ref{E}),
do not mix even and odd $m$ components, and that the rotation term in Eq. (\ref{olla3}) 
vanishes in the limit. In other words, the fore-aft axis of the vesicle,
associated with $R^\smaldu$, and the long ellipsoid axis, 
align at $\pi/4$ with respect to the flow].

\begin{figure}
\begin{center}
\includegraphics[draft=false,width=13.cm]{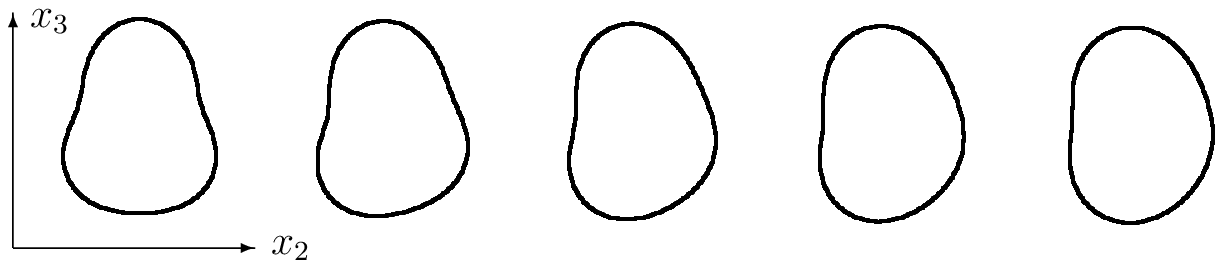}
\caption{Sketch of the vesicle orientation in the $x_2x_3$ plane for $\lambda/\lambda_{cr}=1$
and different values of $\Ca\,\lambda_{cr}$.
From left to right: 
$\Ca\,\lambda_{cr}=0,3.2,6.3,9.5,12.6$.
In all cases, $\hat\kappa<0$, corresponding to drift to $x_2>0$.
The shapes have been drawn starting from Eqs. (\ref{shape},\ref{B1}) and (\ref{olla3}),
greatly exaggerating the amplitude of the components $R^\smalun_{2m}$ and $R^\smaldu_{3m}$.
}
\label{eggfig4}
\end{center}
\end{figure}

For the same reason, 
the fore-aft and the long ellipsoid axis will align for
$\Ca\,\lambda_{cr}\to 0$,
and $\lambda/\lambda_{cr}$ fixed. 
At $\lambda=\lambda_{cr}$, this will occur
along the flow direction, which maximizes $\tilde U^{migr}$ (in fact, $R^\smaldu_{3m}$ has only
$m=0,2$ components). The shift in the vesicle orientation, occurring in the process,
%when $\Ca\,\lambda_{cr}$ is sent to zero for $\lambda/\lambda_{cr}$ fixed,  
is illustrated in Fig. \ref{eggfig4}.

\section{Conclusion}
\label{sec6}
A microswimmer could in principle exploit velocity gradients in the suspending 
fluid  as a means for propulsion. We have seen that a vesicle could obtain this result
by control over the bending rigidity of its membrane. The additional stresses induced
in the membrane by its own inhomogeneities \cite[]{goulian93} could be exploited to 
counteract the effect of the external flow and to allow vesicle shapes and orientations
that would otherwise be forbidden.

In the case of an unbounded
linear shear, a vesicle with such capabilities could migrate transverse to the flow, both in 
the shear plane and perpendicular to it. 
Propulsion is achieved through modulation of
tank-treading by inhomogeneity of the membrane properties, which results in a
constant fore-aft asymmetric shape 
for the vesicle 
(an egg-shape with tip along the flow, in case of in-plane drift) 
and a dipole 
component in velocity perturbation around it, associated with the presence of
a transverse force.

These calculations can be easily extended to the case of a shear flow bounded by a wall,
thus allowing to consider regimes in which a tank-treading vesicle could be made to 
drift towards a wall, rather than away from it, as occurring normally 
with an homogeneous membrane \cite[]{abkarian02}.

We have shown that such migration behaviors (at least, those confined to the shear plane)
could be achieved simply by a stiffening (or softening) of the membrane, 
in response to variations of some local fluid properties, e.g. a chemical concentration.
With an appropriate choice of parameters in the membrane response, a vesicle could be
made to migrate automatically, up or down a concentration gradient in the shear plane,
without the need of 
%internal energy sources and 
complicated control systems.

The results discussed have been obtained through a perturbative calculation in the 
case of a quasi-spherical vesicle. In this regime, the migration velocity is 
smaller than the velocity scale $\alpha R_0$ (the velocity difference
in the shear flow at the vesicle scale), by a factor $\tilde\kappa\epsilon^{1/2}$,
with $\epsilon$ the normalized excess area and $\tilde\kappa$ the degree of
inhomogeneity of the bending energy [see Eqs. (\ref{area}) and (\ref{kappa})].  
However, as the mechanisms for stress generation
by membrane inhomogeneity, and for migration by asimmetry of the vesicle shape,
are essentially geometric, the same behaviors are expected to 
hold also for strongly non-spherical vesicles and strong inhomogeneity in the
bending rigidity.

In the present analysis, no attention has been given to the problem of the
material  that could be utilized to synthesize a membrane with the properties 
described in Eqs. (\ref{kappas},\ref{bound}), or in Eq. (\ref{stiffening}).
In principle, the required local modifications of the membrane could be achieved
by some sort of absorption process or chemical process with the bulk, in 
analogy with what is done in experiments on the Marangoni effect \cite[]{kitahata02}
and osmophoresis \cite[]{nardi99}. It remains to be determined, however, which 
material could satisfy the requirements imposed by Eq. (\ref{stiffening}), 
in particular, slow relaxation of membrane properties, compared to hydrodynamic 
time-scales.
%and some saturation mechanism acting selectively on the homogeneous component 
%of the bending rigidity.
At the present stage, therefore, as in the case of most microswimmer designs, 
also the present vescicle-based one remains at the level of a purely theoretical 
model.

It must be said that the 
adoption of a vesicle based microswimmer design, 
rather than a microcapsule based one, has been motivated
purely by simplicity considerations on the constitutive
law for the membrane.  
After all, tank-treading behaviors, similar to those of vesicles, are observed also 
in the case of microcapsules \cite[][]{barthes80,pozrikidis03,skotheim07}. 
This could open additional possibilities as regards the issue of 
practical realization.
An interesting question is whether a microswimmer design, utilizing an elastic,
rather than fluid membrane, would be more effective in converting fluid stresses
into migration.

\begin{acknowledgments} The author wishes to thank Alexander Farutin and Chaoqui Misbah
for interesting and helpful discussion.
\end{acknowledgments}

%Thus, although probably there are no physical reasons to prevent such
%materials to become available in the feature, the practical realization 
%of the proposed design remain hypothetical. 
%to design microswimmers with the properties described in the present paper, using
%a microcapsule rather than a vesicle structure as a starting point. 

\appendix
\section{Lamb representation}
The Stokes equation for an incompressible fluid are
\beq
\eta\nabla^2\v=\nabla P,
\qquad
\nabla\cdot\v=0,
\label{Stokes}
\eeq
where $P$ is the pressure.
This equation can be solved in the basis of Eq. (\ref{vector_spherical}), 
with boundary conditions $\v=\V$ imposed on a surface $r=R_0$; the result is the
so called Lamb representation of the Stokes equation \cite[]{happel}. 
The solutions for $r<R_0$ and $r>R_0$ read respectively:
\beq
\begin{array}{ll}
v^{in}_{{\rm S}lm}=
\frac{1}{2}\Big[(l+3-(l+1)y^2)V_{{\rm S}lm}
+\sqrt{l(l+1)}(y^2-1)V_{{\rm E}lm}\Big]\,
y^{l-1}\\
v^{in}_{{\rm E}lm}=
\frac{1}{2}\Big[(l+3)\sqrt{\frac{l+1}{l}}(1-y^2)V_{{\rm S}lm}
+(-(l+1)+(l+3)y^2)V_{{\rm E}lm}\Big]\, y^{l-1}\\
v^{in}_{{\rm M}lm}=V_{{\rm M}lm}\, y^l.
\end{array}
\label{inner}
\eeq
and
\beq
\begin{array}{ll}
v^{out}_{{\rm S}lm}=
\frac{1}{2}\Big[(l+(2-l)y^{-2})V_{{\rm S}lm}
+\sqrt{l(l+1)}(l-y^{-2})V_{{\rm E}lm}\Big]\,
y^{-l}
\\
v^{out}_{{\rm E}lm}=
\frac{1}{2}\Big[(2-l)\sqrt{\frac{l}{l+1}}(1-y^{-2})V_{{\rm S}lm}
+(2-l+ly^{-2})V_{{\rm E}lm}\Big]\, y^{-l},
\\
v^{out}_{{\rm M}lm}=V_{{\rm M}lm}\, y^{-1-l},
\end{array}
\label{outer}
\eeq
where  $y=r/R_0$.

The shear flow $\bar\u=\alpha x_2\hat\x_3$ and $\u$, that is the flow inside the vesicle,
are both in the form of $\v^{in}$, while the perturbation $\hat\u$ is in the form
of $\v^{out}$. The lowest order solutions $\hat\u^\smalze$ and $\u^\smalze$ are given
by Eqs. (\ref{outer}) and (\ref{inner}) with $\V=\hat\U^\smalze$ and $\V=\U^\smalze$.
To higher order in $\tilde R$, different harmonics in $\hat\U,\U$ and $\hat\u,\u$ get
mixed due to deviation from spherical shape.

The vector spherical components on the surface $r=R_0$ for the shear flow $\bar\u$ are 
obtained from  Eq. (\ref{inner}):
\beq
\frac{\bar U^\smalze_{{\rm S}2,\pm 1}}{\alpha R_0}=\im\sqrt{\frac{2\pi}{15}},
\qquad
\frac{\bar U^\smalze_{{\rm E}2,\pm 1}}{\alpha R_0}=\im\sqrt{\frac{\pi}{5}}
\quad
{\rm and}
\quad
\frac{\bar U^\smalze_{{\rm M}1,\pm 1}}{\alpha R_0}=\pm\sqrt{\frac{\pi}{3}}.
\label{shear_component}
\eeq
It is easy to see that $U^\smalze_{{\rm M}1,\pm 1}$ is responsible for the vorticity part
of the velocity $\bar\u^{rot}=\frac{1}{2}\alpha(x_2\hat\x_3-x_3\hat\x_2)$.

The force density on the surface $r=R_0$, produced by the internal field in Eq. (\ref{inner}) 
will read, in dimensionless form:
\beq
\begin{array}{l}
g_{{\rm S}lm}(\V)=\frac{1}{\alpha R_0}\Big[
-\frac{2l^2+l+3}{l}V_{{\rm S}lm}+3\sqrt{\frac{l+1}{l}}V_{{\rm E}lm}\Big],
\\
g_{{\rm E}lm}(\V)=\frac{1}{\alpha R_0}\Big[
3\sqrt{\frac{l+1}{l}}V_{{\rm S}lm}-(2l+1)V_{{\rm E}lm}\Big],
\\
g_{{\rm M}lm}(\V)=-\frac{(l-1)}{\alpha R_0}V_{{\rm M}lm}
\end{array}
\label{force_inner}
\eeq
and the one from the external field, Eq. (\ref{outer}):
\beq
\begin{array}{l}
\hat g_{{\rm S}lm}(\V)=
-\frac{1}{\alpha R_0}\Big[\frac{2l^2+3l+4}{l+1}V_{{\rm S}lm}+3\sqrt{\frac{l}{l+1}}
V_{{\rm E}lm}\Big],
\\
\hat g_{{\rm E}lm}(\V)=
\frac{1}{\alpha R_0}\Big[
3\sqrt{\frac{l}{l+1}}V_{{\rm S}lm}-(2l+1)V_{{\rm E}lm}\Big],
\\
\hat g_{{\rm M}lm}(\V)=-\frac{(l+2)}{\alpha R_0}V_{{\rm M}lm}.
\end{array}
\label{force_outer}
\eeq
The total hydrodynamic force $\F$ on the vesicle is obtained from the 
$\mu={\rm S,E}$, $l=1$ components of the outer Lamb solution (\ref{outer}) 
$\hat\u$.
In particular, for the force components along $\hat\x_{1,2}$:
\begin{equation}
\begin{array}{l}
F_1=2\sqrt{6\pi}\eta_{ext}R_0\,Re(\hat U^\smalun_{{\rm S}11}
+\sqrt{2}\hat U^\smalun_{{\rm E}11}),
\\
F_2=-2\sqrt{6\pi}\eta_{ext}R_0\,Im(\hat U^\smalun_{{\rm S}11}
+\sqrt{2}\hat U^\smalun_{{\rm E}11}),
\end{array}
\label{cacca}
\end{equation}
where we have used the fact that, from Eqs. (\ref{boundary_condition_0}) and
(\ref{shear_component}):
$\hat U^\smalze_{\mu 11}=0$ for $\mu={\rm S,E}$. In the absence of external forces,
this is cancelled by the drag force $-D_0\U^{drift}$ on a vesicle migrating with 
velocity $\U^{drift}$, where, to lowest order in $\epsilon$, $D_0$ is the Stokes drag
by a spherical vesicle.
Now, a spherical vesicle with an inextensible membrane
will behave with respect to the fluid as a rigid object [except for solenoidal flow components 
$\U_{{\rm M}lm}$ on the surface, that do not couple with $\U^{drift}$; see Eqs. 
(\ref{U1}) and (\ref{inextensibility})]. Hence, $D_0$ is the drag coefficient
of a rigid sphere. 
The drift velocity of the vesicle will be therefore, to lowest order in $\epsilon$:
\beq
\U^{drift}=\F/D_0,
\label{drift}
\eeq
where $D_0=6\pi\eta_{ext}R_0$ is the Stokes drag for a rigid sphere of radius $R_0$.

%\beq
%\begin{array}{l}
%U_1^{drift}=\frac{2}{\sqrt{6\pi}}Re(\hat u^\smalun_{{\rm S}11}+
%\sqrt{2}\hat u^\smalun_{{\rm E}11})_{r=R_0},
%\\
%U_2^{drift}=-\frac{2}{\sqrt{6\pi}}Im(\hat u^\smalun_{{\rm S}11}+
%\sqrt{2}\hat u^\smalun_{{\rm E}11})_{r=R_0}.
%\label{drift}
%\end{array}
%\eeq
%

\section{Dynamics of the homogeneous membrane in a shear flow}
Assuming a tank-treading regime, the shape of a vesicle in the shear flow (\ref{shear})
is obtained from Eq. (\ref{olla2}) setting the time derivative equal to zero.
Using Eqs. (\ref{ABCD},\ref{Omega}) and (\ref{C}), and setting to zero the inhomogeneous
contribution $D_l\tilde\kappa_{lm}$:
\beq
&&\frac{\im \Ca\,\Lambda}{2}\tilde R^\smalun_{21}+B_2\tilde R^\smalun_{22}=0,
\nonumber
\\
&&\im \Ca\,\Lambda\Big[\frac{\tilde R^\smalun_{22}}{2}+\frac{\sqrt{6}}{4}\tilde R^\smalun_{20}\Big]+B_2\tilde R^\smalun_{21}
=2\im \Ca\,\sqrt{\frac{10\pi}{3}},
\\
&&\frac{\im\sqrt{6} \Ca\,\Lambda}{2}\tilde R^\smalun_{21}+B_2\tilde R^\smalun_{20}=0,
\qquad \Lambda=A_2\lambda,
\nonumber
\eeq
which gives the result
\beq
\tilde R^\smalun_{20}=K \Ca\,\Lambda\sqrt{6},
\quad
\tilde R^\smalun_{21}=2KB_2\im,
\quad
\tilde R^\smalun_{22}=K \Ca\,\Lambda,
\label{B1}
\eeq
where
\beq
K=\frac{ \Ca}{B_2^2+( \Ca\,\Lambda)^2}\sqrt{\frac{10\pi}{3}}.
\label{K}
\eeq
Substituting Eqs. (\ref{B1}) and (\ref{K}) into the area constrain (\ref{area}),
we obtain 
\beq
B_2=\sqrt{\frac{160\pi \Ca^2}{3\epsilon}-( \Ca\,\Lambda)^2},
\label{B2}
\eeq
from which we obtain Eq. (\ref{barT}).
Exploiting the first of Eq. (\ref{ABCD}), we see that tank-treading is possible for 
\beq
\lambda<\lambda_{cr}=\frac{24}{23}\sqrt{\frac{10\pi}{3\epsilon}},
\nonumber
\eeq
that coincides, to leading order in $\epsilon$, with
the result in \cite{farutin10}.
%which is an appropriate expression for the transition out of the tank-treading
%regime, only for a quasi-spherical vesicle.
Substituting Eq. (\ref{B2}) into Eqs. (\ref{K}) and (\ref{B1}), we obtain Eq. (\ref{R1}).

\section{Matrix elements involving scalar and vector spherical harmonics}
We provide below expressions for the matrix elements entering Eqs. 
(\ref{kappa1},\ref{olla3},\ref{E}).
\beq
&&\langle 30|Y_{21}\sin\theta\sin\phi\rangle=\im\sqrt{\frac{3}{70}},
\qquad\qquad\quad
\langle 31|Y_{20}\sin\theta\sin\phi\rangle=\im\sqrt{\frac{3}{35}},
\nonumber
\\
&&\langle 31|Y_{22}\sin\theta\sin\phi\rangle=\frac{\im}{\sqrt{70}},
\qquad\qquad\quad \ \,
\langle 32|Y_{21}\sin\theta\sin\phi\rangle=\frac{\im}{\sqrt{7}},
\nonumber
\\
&&\langle 33|Y_{22}\sin\theta\sin\phi\rangle=\im\sqrt{\frac{3}{14}},
\qquad\qquad\quad
\langle 30|\Y_{{\rm E}20}\cdot\tilde\nabla Y_{10}\rangle
=-3\sqrt{\frac{3}{35\pi}},
\nonumber
\\
&&\langle 31|\Y_{{\rm E}21}\cdot\tilde\nabla Y_{10}\rangle
=-2\sqrt{\frac{6}{35\pi}},
\qquad\quad\ \ \ 
\langle 32|\Y_{{\rm E}22}\cdot\tilde\nabla Y_{10}\rangle
=-\sqrt{\frac{3}{7\pi}},
\nonumber
\\
&&\langle 30|Y_{10}|20\rangle=\frac{3}{2}\sqrt{\frac{3}{35\pi}},
\qquad\qquad\qquad\ \ \ 
\langle 31|Y_{10}|21\rangle=\sqrt{\frac{6}{35\pi}},
\nonumber
\\
&&\langle 32|Y_{10}|22\rangle=\frac{1}{2}\sqrt{\frac{3}{7\pi}},
\qquad\qquad\qquad\quad\ \,
\langle{\rm E}30|\tilde\nabla Y_{10}|20\rangle=\frac{3}{\sqrt{35\pi}},
\nonumber
\\
&&\langle{\rm E}31|\tilde\nabla Y_{10}|21\rangle=2\sqrt{\frac{2}{35\pi}},
\qquad\qquad\qquad
\langle{\rm E}32|\tilde\nabla Y_{10}|22\rangle=\frac{1}{\sqrt{7\pi}}.
\nonumber
\eeq

\end{document}